\documentclass{article}

\usepackage{arxiv}

\usepackage[utf8]{inputenc} 
\usepackage[T1]{fontenc}    
\usepackage{hyperref}       
\usepackage{url}            
\usepackage{booktabs}       
\usepackage{amsfonts}       
\usepackage{nicefrac}       
\usepackage{microtype}      
\usepackage{lipsum}		
\usepackage{graphicx}
\usepackage{natbib}
\usepackage{doi}
\usepackage{amsmath}
\usepackage{subcaption}

\title{SpikGPT: A High-Accuracy and Interpretable Spiking Attention Framework for Single-Cell Annotation}

\author{
  Min Huang \\
  Department of Biomedical Informatics \\
  Emory University \\
  Atlanta, GA\\
   \And
  Rishikesan Kamaleswaran \\
  Department of Surgery \\
  Duke University \\
  Durham, NC\\
}

\hypersetup{
pdftitle={SpikGPT},
pdfsubject={q-bio.NC, q-bio.QM},
pdfauthor={Min Huang, Rishikesan Kamaleswaran},
pdfkeywords={scRNA-seq annotation, Spiking attention, transformer},
}

\makeatletter
\providecommand\phantomsubcaption{%
  \refstepcounter\@captype
}
\makeatother

\begin{document}
\maketitle

\begin{abstract}
	Accurate and scalable cell type annotation remains a challenge in single-cell transcriptomics, especially when datasets exhibit strong batch effects or contain previously unseen cell populations. Here we introduce SpikGPT, a hybrid deep learning framework that integrates scGPT-derived cell embeddings with a spiking Transformer architecture to achieve efficient and robust annotation. scGPT provides biologically informed dense representations of each cell, which are further processed by a multi-head Spiking Self-Attention mechanism, energy-efficient feature extraction. Across multiple benchmark datasets, SpikGPT consistently matches or exceeds the performance of leading annotation tools. Notably, SpikGPT uniquely identifies unseen cell types by assigning low-confidence predictions to an “Unknown” category, allowing accurate rejection of cell states absent from the training reference. Together, these results demonstrate that SpikGPT is a versatile and reliable annotation tool capable of generalizing across datasets, resolving complex cellular heterogeneity, and facilitating discovery of novel or disease-associated cell populations.
\end{abstract}


\section{Introduction}


Single-cell RNA sequencing (scRNA-seq) has powerful ability to dissect cellular heterogeneity, enabling the profiling of gene expression at single-cell resolution across diverse tissues and conditions\cite{plass2018cell}\cite{cao2019single}\cite{kim2018chemoresistance}. A critical step in scRNA-seq analysis is cell-type annotation by clustering, which assigns biological meaning to expression profiles and lays the foundation for downstream interpretation and discovery\cite{butler2018integrating}\cite{stuart2019comprehensive}. In traditional way, researchers label cell types manually based on the clustering and prior knowledge\cite{xie2021automatic}\cite{huang2021evaluation}. However, labor annotation is time consuming and non-reproducible. Therefore, scRNA-seq cell-type annotation tools were developed and achieved superior performance. Despite numerous advances, automated cell-type annotation remains challenging due to the high dimensionality, sparsity, and noise inherent in scRNA-seq data. Conventional machine learning methods often struggle to generalize across datasets\cite{aran2019reference}\cite{pliner2019supervised}\cite{kiselev2018scmap}\cite{alquicira2019scpred}, while deep learning models can be computationally intensive and lack biological interpretability\cite{xu2021probabilistic}\cite{ma2020actinn}. Recent transformer-based approaches have shown promise by learning rich embeddings from scRNA-seq data\cite{chen2023transformer}\cite{xu2023ciform}. However, these models are not optimized for sparse biological signals and often require extensive computational resources for training and inference.

To address these limitations, we developed a novel cell-type annotation framework called SpikGPT that integrates the strengths of biological language modeling with biologically inspired attention mechanisms. Our model is predicated on the paradigm of biological language models. This approach treats gene expression profiles as structured sequences, analogous to sentences in natural language, thereby facilitating the learning of context-aware representations that intrinsically capture the co-expression relationships and underlying regulatory logic among genes. By leveraging this representation paradigm, we utilize scGPT-generated\cite{cui2024scgpt} embeddings to encode rich semantic information from single-cell gene expression data. To further enhance computational efficiency and biological plausibility, Spiking Self-Attention (SSA) mechanism that mimics the sparse, event-driven nature of neuronal firing were introduced. Our multi-head spiking attention module processes the scGPT embeddings through dynamic spiking neuron models, enabling sparse, energy-efficient, and interpretable information flow. This hybrid architecture effectively combines the sequence modeling power of language models with biologically grounded sparse computation, yielding a model that is both accurate and computationally lightweight.

We demonstrate that our approach, SpikeGPT, significantly outperforms existing annotation tools in both accuracy and computational efficiency across multiple benchmark scRNA-seq datasets. Crucially, SpikeGPT exhibits superior generalization capability for predictions on external and heterogeneous datasets, affirming its robustness. Furthermore, by learning deep, universal representations of cellular states, our model possesses the unique ability to accurately predict and characterize novel or 'unseen' cell types not present in the training data, overcoming a major limitation of conventional methods. By grounding its design in biological language and neural computation principles, SpikeGPT offers a powerful, highly generalizable, and interpretable solution for large-scale single-cell analysis.

\section{Method}




\subsection*{SpikGPT model}
We propose SpikGPT, which incorporates the scGPT embedding into Spiking self attention. 
For each cell, gene expression values from the scRNA-seq matrix are first embedded into a dense vector using scGPT, a transformer-based model pretrained on large-scale single-cell datasets. This embedding captures the global transcriptional profile of the cell and serves as input to our annotation framework. 
\begin{equation}
E = scGPT(X)
\label{eq:scGPT}
\end{equation}
Where $X \in \mathbb{R}^{n \times g}$ is the original scRNA-seq matrix, with $n$ cells and $g$ genes, $E \in \mathbb{R}^{n \times k}$ is the resulting dense embedding matrix, where each cell is represented by a k-dimensional feature vector.




After obtaining the cell embedding matrix $E \in \mathbb{R}^{n \times k}$ from scGPT, we apply two expansion steps to adapt the data for input into the Spikformer model and to enhance its capacity for robust and expressive learning.

First, each individual cell embedding $e_i \in \mathbb{R}^{k}$ is repeated $m$ times along a new axis, resulting in a tensor $E' \in \mathbb{R}^{n \times m \times k}$, where m is a hyper-parameter that can be manually set, with a default of 300. This expansion allows the model to construct multiple parallel feature representations for each cell, increasing the richness and flexibility of representation. Although the repeated vectors are initially identical, subsequent non-linear transformations (such as spiking activations and attention mechanisms) introduce diversity and enable independent adaptation across feature channels.

Second, the entire $E'$ matrix is repeated $T$ times along a new dimension, forming the final input tensor $E'' \in \mathbb{R}^{T \times n \times m \times k}$. $T$ is a hyper-parameter that can be manually set, with a default of 4. This dimension can be interpreted as multiple stochastic views of the same dataset, promoting robustness and generalization by allowing the model to learn invariant features across repeated input exposures.

Together, these two expansion steps enhance the model's ability to represent complex cellular states and improve training stability under sparse and spiking activation dynamics.




The final input tensor $E^{\prime\prime} \in \mathbb{R}^{T \times n \times m \times k}$, obtained through the aforementioned embedding expansion, is then fed into the Spiking Self-Attention (SSA) module, which is organized into $H$ parallel attention heads.

For each attention head $i$ $(i = 1, 2, \dots, H)$, the Query ($Q_i$), Key ($K_i$), and Value ($V_i$) matrices are computed using separate learnable linear projections:
\begin{equation}
Q_i = E^{\prime\prime} W_Q^{(i)},\quad K_i = E^{\prime\prime} W_K^{(i)},\quad V_i = E^{\prime\prime} W_V^{(i)}
\label{eq:qkv_heads}
\end{equation}

where $W_Q^{(i)}, W_K^{(i)}, W_V^{(i)} \in \mathbb{R}^{k \times d_h}$ are the projection weight matrices for the $i$-th attention head, and $d_h$ denotes the dimensionality of each head. Using independent projections per head allows the model to capture diverse patterns across different subspaces, which helps stabilize training and improve representational capacity.

Next, the projected matrices pass through distinct spiking neuron layers to generate sparse, spike-form representations. To introduce biologically inspired sparse activation into the attention mechanism, we adopt Leaky Integrate-and-Fire (LIF) neurons in all spiking neuron layers of SpikGPT.  Specifically, each Query ($Q_i$), Key ($K_i$), and Value ($V_i$) matrix is passed through a corresponding spike neuron layer—denoted as $\text{SN}_Q$, $\text{SN}_K$, $\text{SN}_V$ —before computing the attention outputs:

\begin{equation}
Q_i' = \text{SN}_Q(Q_i),\quad K_i' = \text{SN}_K(K_i),\quad V_i' = \text{SN}_V(V+i)
\label{eq:spike_neuron}
\end{equation}


The LIF neurons are implemented using the SpikingJelly framework. For parameters of it, the time constant $\tau=2.0$ controls the rate of membrane potential decay. The step\_mode='m' setting enables multi-step (temporal) mode, allowing the neuron to integrate inputs across multiple discrete time steps. All spikes are binary, temporally sparse events, enhancing both computational efficiency and model generalizability.



The spiking self-attention operation is then computed as:
\begin{equation}
\text{SSA}_i = \text{SN}(Q_i' K_i'^T V_i')
\label{eq:SSA}
\end{equation}

Here, $\text{SN}$ denotes the spiking neuron function applied to the output of each attention head. For each of the $H$ attention heads, the query ($Q_i'$), key ($K_i'$), and value ($V_i'$) matrices are separately projected and then passed through the attention mechanism followed by a spiking neuron layer. 



The outputs from all attention heads are then concatenated and passed through a multi-layer perceptron (MLP) to produce the final feature representation:
\begin{equation}
O = \text{MLP} \left( \text{concat}(\text{SSA}_1, \dots, \text{SSA}_H) \right)
\label{eq:O}
\end{equation}

Here, $O$ denotes the refined feature matrix obtained after merging the outputs of all attention heads. The concatenation step aggregates information from different subspaces, while the MLP transforms the combined features into a more expressive latent representation.

Afterward, the feature matrix \( O \) is globally averaged across spatial dimensions, reducing it to a compact, cell-level feature vector. Finally, the condensed feature vector is passed through a fully connected classification head:
\begin{equation}
\hat{y} = \text{softmax}(W_c O + b_c)
\label{eq:softmax}
\end{equation}

where $W_c$ is a learnable weight matrix mapping the extracted features to the desired number of cell types, and $b_c$ is the corresponding bias vector. Here, $\hat{y} \in \mathbb{R}^{C}$ denotes the predicted probability distribution over $C$ cell types.
The final predicted label is determined by taking the class with the highest probability $\text{argmax}(\hat{y})$, and mapping it to the corresponding cell type name from the predefined label set.

\subsection*{Model training}
The reference dataset was randomly split into 70\% for training and 30\% for testing. Model performance was evaluated using accuracy, defined as the ratio of correctly predicted samples to the total number of samples. The loss was computed using the cross-entropy loss function.


Training was conducted using stochastic gradient descent (SGD) with default hyperparameters: an initial learning rate of 0.01, momentum of 0.9, and weight decay of 1e-4. A cosine learning rate decay schedule was applied to prevent large parameter updates in the later stages of training. The model was trained for 10 epochs by default, with a batch size of 128.

To mitigate overfitting, a dropout rate of 0.1 was applied to the MLP layers during training. All hyperparameters, including dropout, are configurable and can be adjusted by users depending on the dataset or application scenario.

\subsection*{Data preprocess}
scRNA-seq datasets were obtained from public database. The summary is in sup table1. Scanpy (1.9.8) was applied for standard data preprocess, including normalization (counts per cell were normalized to a total of 10,000), log transformation, and feature scaling. These steps ensured that the input data were suitable for downstream scGPT embedding and other benchmark annotation tools.

\subsection*{Other annotation tools}
To evaluate the performance of our model, we conducted benchmark analysis against eight widely used cell-type annotation tools: scGPT\cite{cui2024scgpt}, SingleR\cite{aran2019reference}, scANVI\cite{xu2021probabilistic}, CCA (Seurat v5 integration)\cite{stuart2019comprehensive}, Scmap\cite{kiselev2018scmap}, Garnett\cite{pliner2019supervised}, SciBet\cite{li2020scibet}, and scPred\cite{alquicira2019scpred}. 
All models were evaluated on the same reference and query datasets to ensure fair comparison. Each method was run using default parameters or following best practices as recommended in their respective publications. To evaluate the performance of all benchmark methods, accuracy was calculated as the proportion of correctly predicted cell types relative to the ground-truth labels, as well as precision, recall and F1-score.

\section{Result}

\subsection{The structure of SpikGPT}

\begin{figure}[htbp]
  \centering
  \includegraphics[width=1\textwidth]{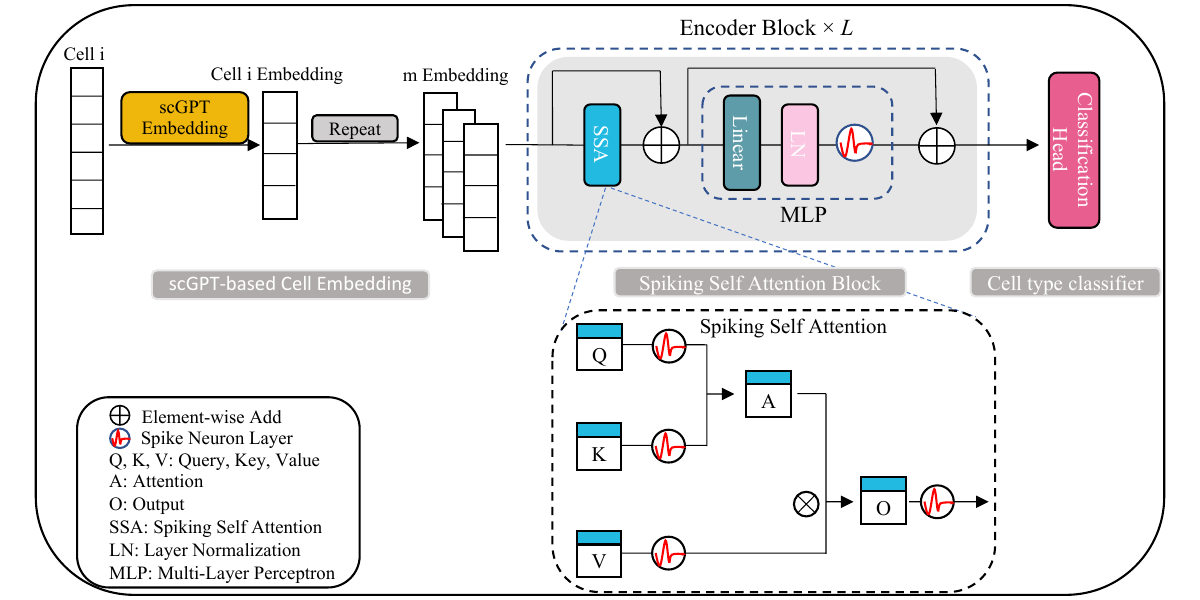}
  \caption{The model is trained using single-cell RNA sequencing (scRNA-seq) data and corresponding cell-type labels. First, gene expression profiles are transformed into contextualized embeddings using a pre-trained scGPT module, producing tokenized representations of individual cells. These tokens are then used to compute query (Q), key (K), and value (V) matrices via learnable linear projections. Each matrix is passed through a corresponding spiking neuron layer to produce sparse, temporally encoded sequences. Spiking Self-Attention is then applied, where attention weights are computed based on the compatibility between Q and K, and used to weight the corresponding V to generate the output (O). The output is further processed by an additional spiking activation and a multi-layer perceptron (MLP) with normalization. In the Multi-Head Spiking Self-Attention module, the attention mechanism is applied in parallel across multiple heads (H), and the resulting outputs are concatenated and integrated. The final output O is averaged across spatial dimensions to produce a compact cell-level latent representation. This vector is used as input to a downstream fully connected classifier for cell-type prediction.}
  \label{fig:structure}
\end{figure}

This model is a high-performance single-cell RNA-seq annotation tool that combines scGPT-based embeddings with a Spiking Self-Attention (SSA) mechanism. This design allows the model to effectively capture the complex, sparse relationships inherent in single-cell gene expression data, resulting in more accurate and interpretable cell-type annotation. The model consists of three main components: scGPT-based Cell Embedding, Spiking Self-Attention, and Cell-Type Classifier (Fig~\ref{fig:structure}).

The first component, the Cell Embedding, uses scGPT to transform high-dimensional gene expression profiles into dense, context-aware embeddings. This step effectively compresses the noisy, sparse gene expression space into a lower-dimensional, biologically meaningful feature space, providing a strong foundation for accurate cell-type annotation. To enhance the model’s ability to capture diverse biological signals, each cell embedding is repeated multiple times, forming a richer, multi-token representation for each cell, which is critical for capturing subtle differences between closely related cell types.


The Spiking Self-Attention, which forms the core of the model, introduces a novel attention mechanism inspired by biological spiking neurons. Unlike traditional self-attention relying on dense matrix multiplications, this mechanism employs spiking neuron layers that integrate projected input currents over time to update their membrane potentials. When the membrane potential surpasses a fixed firing threshold, a spike is generated and the potential is reset, thereby enabling sparse, energy-efficient activation dynamics. This biologically inspired design allows the layer to process repeated embeddings in parallel through multiple attention heads, capturing diverse aspects of cellular states. Consequently, the model efficiently encodes both short- and long-range dependencies between genes while maintaining computational efficiency and biological interpretability.

Finally, the output of the attention layer is aggregated into a compact, globally pooled feature representation, which is then fed into the Cell-Type Classifier. This final layer maps the refined feature vectors to the target cell types, producing a set of predicted cell-type probabilities. By integrating the outputs of multiple attention heads, this approach effectively captures the complex, context-dependent relationships within the data, enhancing both the accuracy and interpretability of cell-type assignments.

\subsection{SpikGPT Provides Robust and Efficient Annotation of Single-Cell Data}

\begin{figure}[htbp]
  \centering
  \includegraphics[width=0.7\textwidth]{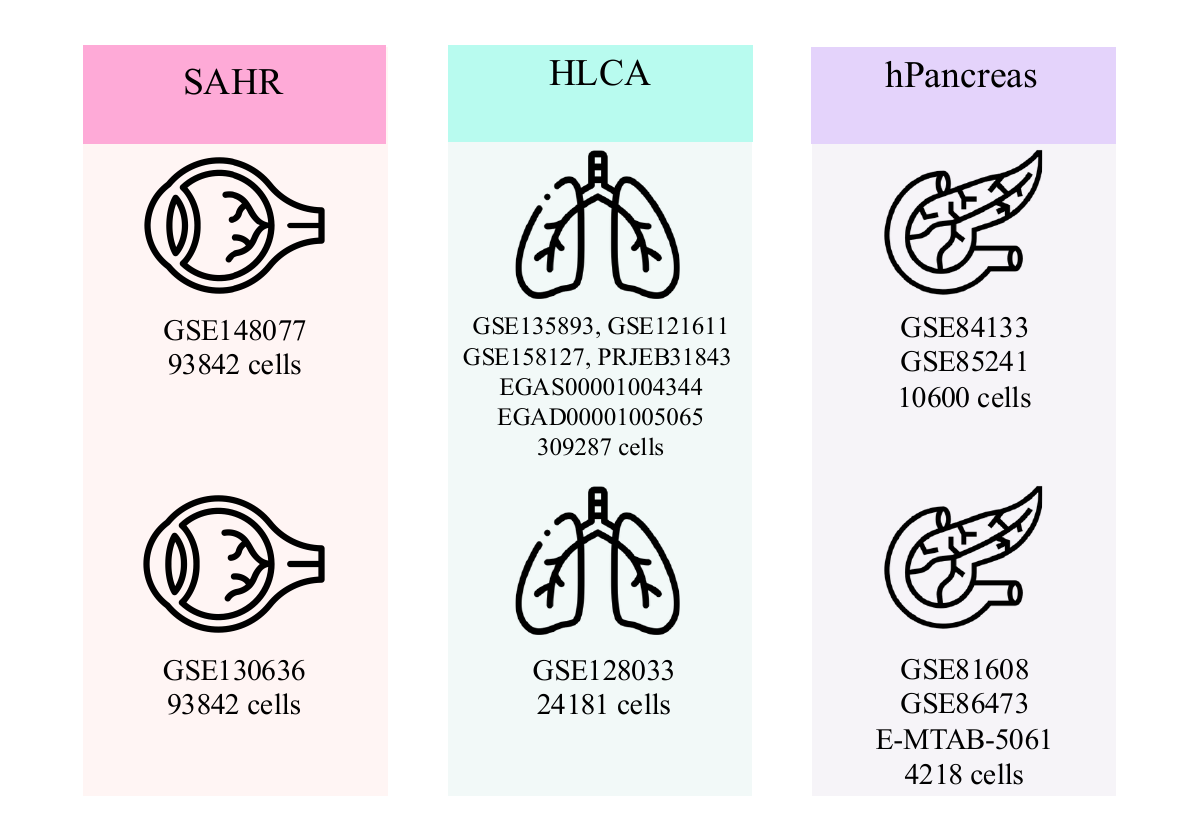}
  \caption{Reference (up) and query (down) datasets in SAHR, HLCA and hPancreas are split by data source.}
  \label{fig:cells}
\end{figure}

We test SpikGPT on 2 different datasets: Single-cell atlas of the human retina (SAHR), The integrated Human Lung Cell Atlas (HLCA) (shown in Fig~\ref{fig:cells}). They all contain “ground truth” cell types from original papers. We also compared SpikGPT with other 8 popular cell type annotators, including foundation models (scGPT), reference-based tools (SingleR, Scmap, CCA), probabilistic approaches (scANVI), and supervised classifiers (Garnett, SingleCellNet, scPred). We compared the accuracy, Precision, Recall and F1-score. SpikGPT consistently achieves competitive performance across all benchmark datasets.

\begin{figure}[htbp]
  \centering
  \includegraphics[width=0.7\textwidth]{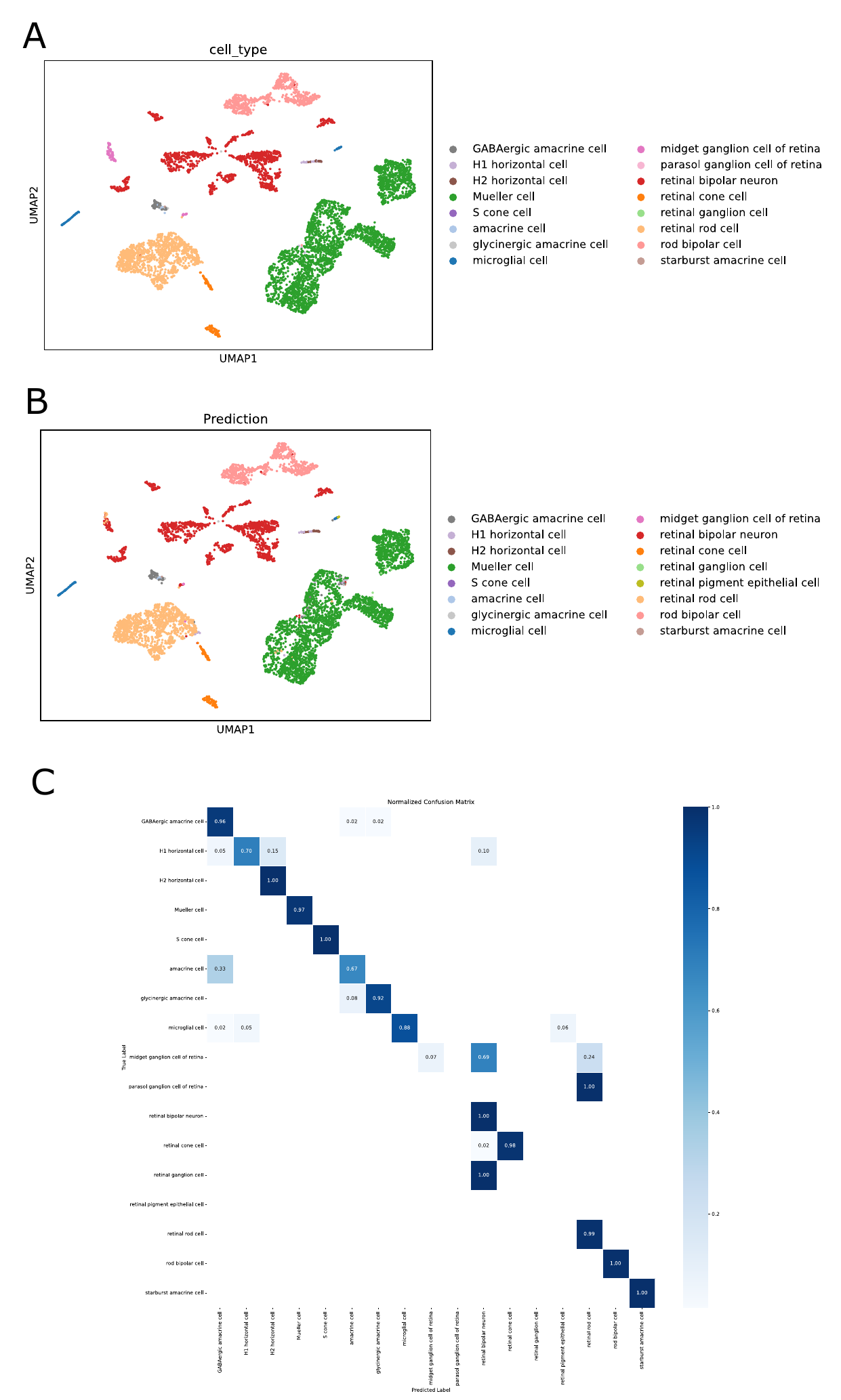}
  \caption{SpikGPT prediction result in the SAHR dataset. A)UMAP visualization of the query dataset colored by ground truth cell-type annotations. B)UMAP of the same dataset with predicted labels from SpikGPT,  C) The confusion matrix displays the distribution of predicted labels (columns) for each ground truth cell type (rows).}
  \phantomsubcaption\label{fig:retinaA}
  \phantomsubcaption\label{fig:retinaB}
  \phantomsubcaption\label{fig:retinaC}
\end{figure}

We first benchmarked on SAHR. We selected GSE148077 (93,842 cells) as reference dataset and GSE130636 (6,061 cells) as query dataset. SpikGPT was trained on GSE148077 and tested on GSE130636. UMAP visualizations of the test dataset, using both ground truth and predicted labels, are shown in Fig~\ref{fig:retinaA} and Fig~\ref{fig:retinaB}. The confusion matrix is presented in Fig~\ref{fig:retinaC}.

Overcall performance is shown in Table~\ref{tab:annotation_comparison_SAHR}. SpikGPT achieved the highest overall accuracy (0.991), outperforming all other methods except for CCA (0.992). However, in macro-level metrics that reflect performance across both abundant and rare cell types, SpikGPT performed more consistently. For example, its macro F1-score (0.713) was competitive with that of scPred (0.827) and Garnett (0.709), but it did so without depending on feature engineering or manual marker input.

Although some methods such as scPred and Garnett achieved higher recall, this came at the cost of lower precision or accuracy. For example, Garnett's recall was 0.854, but its precision was only 0.637. In contrast, SpikGPT provided a more balanced performance across all metrics, demonstrating robustness without the need for marker-based supervision.

\begin{table}[htbp]
\centering
\caption{Performance comparison of cell type annotation methods on dataset SAHR}
\label{tab:annotation_comparison_SAHR}
\begin{tabular}{lcccc}
\toprule
\textbf{Method} & \textbf{Accuracy} & \textbf{Macro Precision} & \textbf{Macro Recall} & \textbf{Macro F1-score} \\
\midrule
SpikGPT         & 0.991 & 0.722  & 0.709  & 0.713  \\
scGPT           & 0.970  & 0.679 & 0.714 & 0.667 \\
SingleR         & 0.557  & 0.450  & 0.641  & 0.444  \\
scANVI          & 0.983 & 0.492 & 0.546 & 0.508 \\
CCA             & 0.992  & 0.797  & 0.660  & 0.894  \\
Scmap-Cluster   & 0.575  & 0.749  & 0.441  & 0.463  \\
Scmap-Cell      & 0.954  & 0.763  & 0.656  & 0.703  \\
Garnett         & 0.928  & 0.637  & 0.854  & 0.709  \\
SingleCellNet   & 0.944  & 0.566  & 0.820  & 0.685  \\
scPred          & 0.982  & 0.787  & 0.741  & 0.827  \\
\bottomrule
\end{tabular}
\end{table}

\begin{figure}[htbp]
  \centering
  \includegraphics[width=0.7\textwidth]{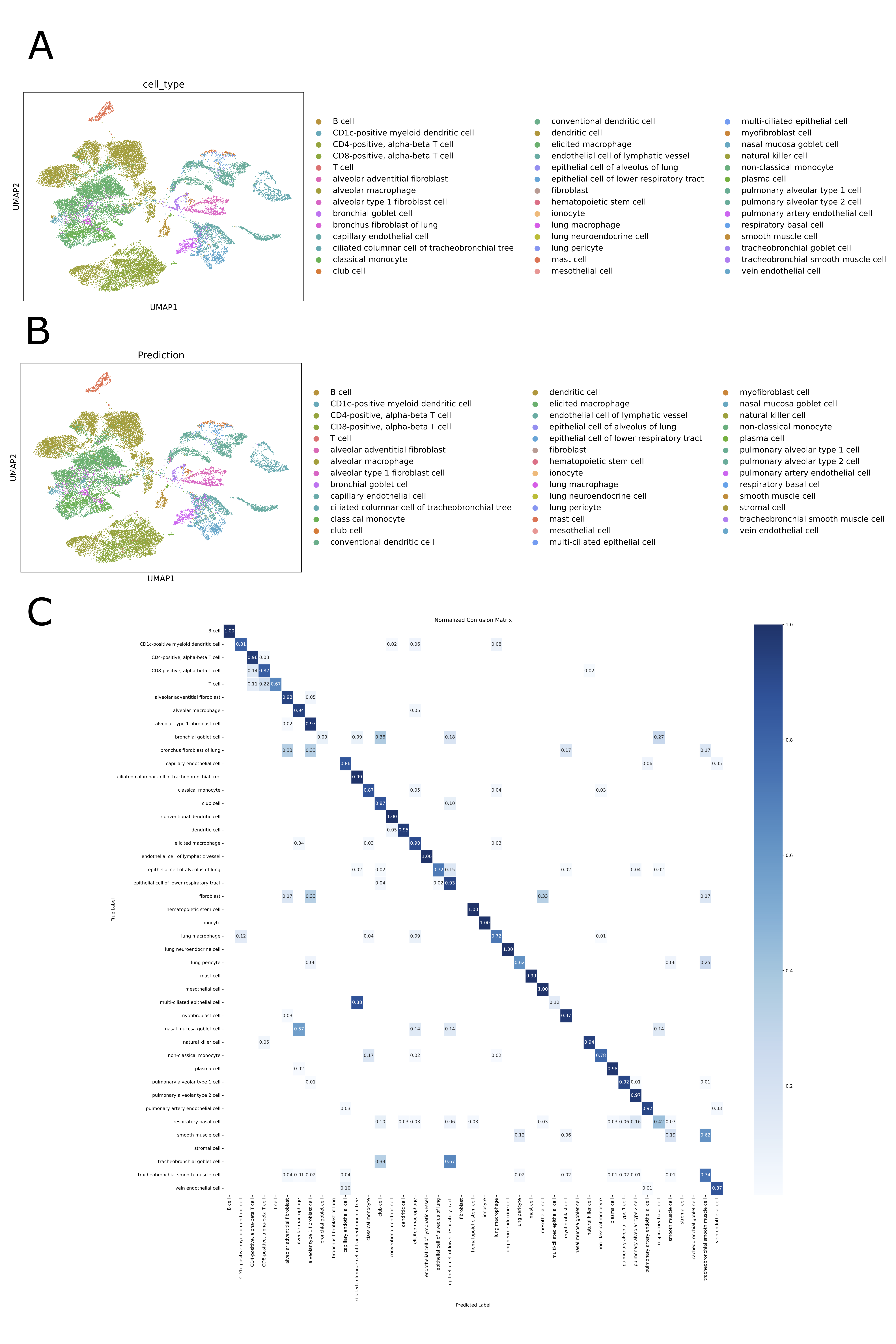}
  \caption{SpikGPT prediction result in the HLCA dataset. A)UMAP visualization of the query dataset colored by ground truth cell-type annotations. B)UMAP of the same dataset with predicted labels from SpikGPT,  C) The confusion matrix displays the distribution of predicted labels (columns) for each ground truth cell type (rows).}
  \phantomsubcaption\label{fig:lungA}
  \phantomsubcaption\label{fig:lungB}
  \phantomsubcaption\label{fig:lungC}
\end{figure}

To further test the model's robustness under realistic data heterogeneity and batch effects, we benchmarked on HLCA. SpikGPT was trained on HLCA core datasets (GSE135893, GSE121611, GSE158127, PRJEB31843, EGAS00001004344, EGAD00001005065)(309,287 cells), and tested on GSE128033 (24,181 cells). Since the reference set includes multiple independent studies, batch effects were present, adding complexity to the task. UMAP visualizations for the test data are shown in Fig~\ref{fig:lungA} and Fig~\ref{fig:lungB}, and confusion matrices in Fig~\ref{fig:lungC}.

Overcall performance is shown in Table~\ref{tab:annotation_comparison_HLCA}. SpikGPT achieved an accuracy of 0.92, macro precision of 0.752, macro recall of 0.721, and macro F1-score of 0.711, making it one of the top-performing methods in this challenging scenario. In comparison, scGPT showed slightly lower accuracy (0.900) and F1-score (0.693), while most traditional methods struggled to handle the data complexity and batch effects.

Among the compared methods, many reference-based approaches (e.g., SingleR, Scmap) exhibited sharp performance drops, especially in macro recall and F1-score, likely due to their sensitivity to batch effects and lack of learned embeddings. scANVI also underperformed in macro metrics, despite its probabilistic modeling. CCA achieved a high macro precision (0.812) but lower recall (0.643), indicating it was more conservative in its predictions.

In contrast, SpikGPT consistently delivered balanced and robust performance, suggesting its ability to generalize well across heterogeneous datasets and capture both abundant and rare cell types without relying on manual annotation or explicit batch correction.

\begin{table}[htbp]
\centering
\caption{Performance comparison of cell type annotation methods on dataset HLCA}
\label{tab:annotation_comparison_HLCA}
\begin{tabular}{lcccc}
\toprule
\textbf{Method} & \textbf{Accuracy} & \textbf{Macro Precision} & \textbf{Macro Recall} & \textbf{Macro F1-score} \\
\midrule
SpikGPT         & 0.920  & 0.752  & 0.721  & 0.711 \\
scGPT           & 0.900  & 0.701 & 0.732  & 0.693 \\
SingleR         & 0.343  & 0.237  & 0.290  & 0.207 \\
scANVI          & 0.733 & 0.279 & 0.371 & 0.297 \\
CCA             & 0.882  & 0.812  & 0.643  & 0.761 \\
Scmap-Cluster   & 0.499  & 0.673  & 0.340  & 0.422 \\
Scmap-Cell      & 0.212  & 0.832  & 0.213  & 0.411 \\
Garnett         & 0.048 & 0.441  & 0.077 & 0.347 \\
SingleCellNet   & 0.732  & 0.630  & 0.601  & 0.626 \\
\bottomrule
\end{tabular}
\end{table}

\subsection{SpikGPT enables discovery of new cell types}

To assess the model’s ability to recognize novel cell populations absent from the training data, we conducted a simulation experiment using the human Pancreas (hPancreas) datasets shown in Fig~\ref{fig:cells}. SpikGPT was trained on GSE84133 and GSE85241 (10600 cells), and test on GSE81608, E-MTAB-5061 and GSE86473 (4218 cells). Specifically, we removed all ‘alpha cells’ from the reference dataset to mimic a realistic scenario in which a biologically meaningful and abundant cell type is missing. The query dataset remained unchanged and included all cell types, including alpha cells.

SpikGPT outputs a probability distribution over all known cell types for each query cell. To enable detection of unseen cell types, we implemented a post hoc rejection mechanism: if the highest predicted probability for a cell was below a predefined threshold (0.95), the cell was labeled as ‘Unknown’. This confidence-based filtering prevents overconfident misclassifications when unfamiliar cell types are encountered.

\begin{figure}[htbp]
  \centering
  \includegraphics[width=1\textwidth]{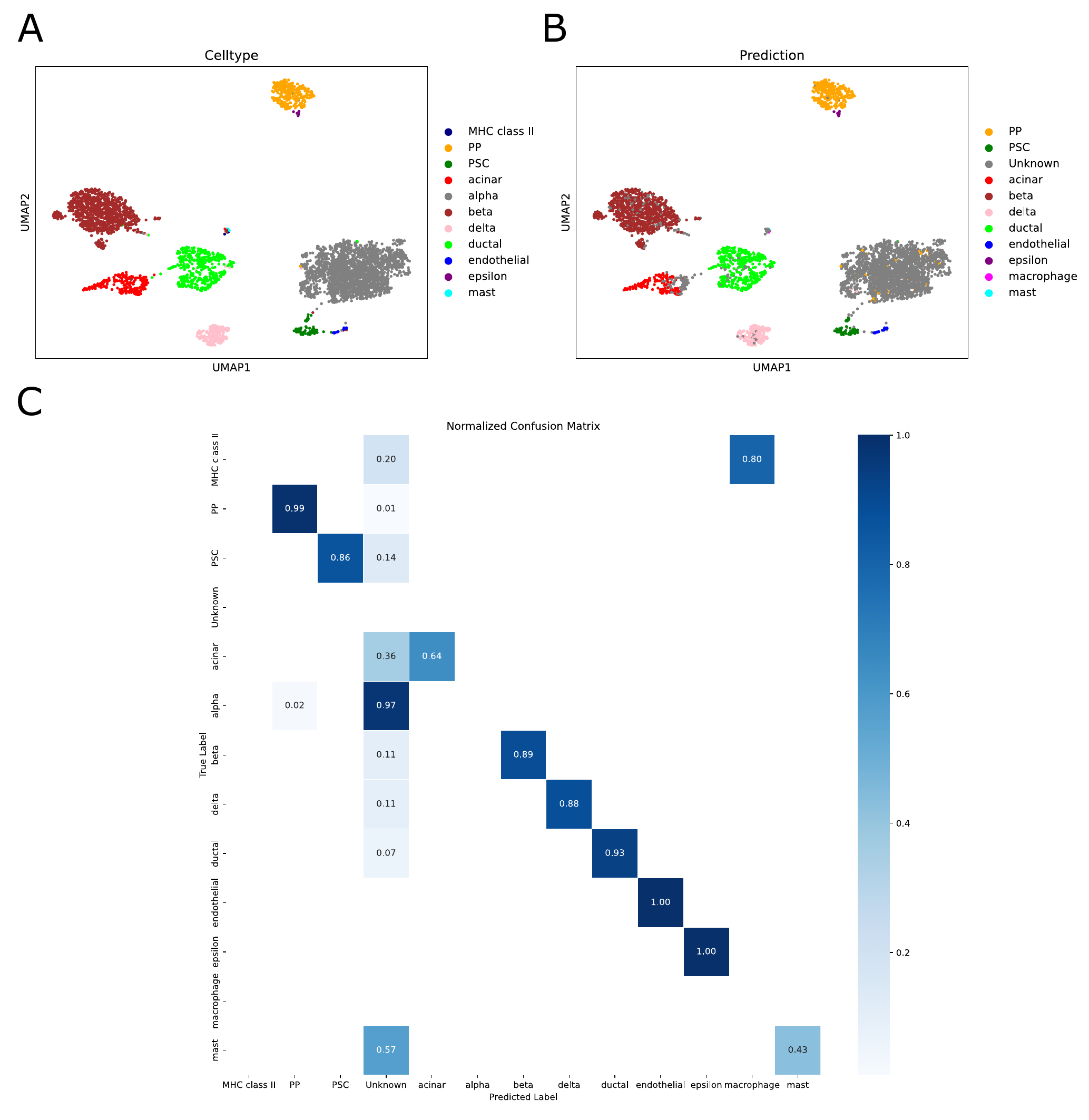}
  \caption{SpikGPT accurately identifies and isolates previously unseen alpha cells as ‘Unknown’ in the hPancreas dataset. A)UMAP visualization of the query dataset colored by ground truth cell-type annotations. B)UMAP of the same dataset with predicted labels from SpikGPT, showing successful separation of alpha cells. C)A large proportion of alpha cells are labeled as ‘Unknown’, with minimal misclassification of other cell types. The confusion matrix displays the distribution of predicted labels (columns) for each ground truth cell type (rows).}
  \phantomsubcaption\label{fig:hPancreasA}
  \phantomsubcaption\label{fig:hPancreasB}
  \phantomsubcaption\label{fig:hPancreasC}
\end{figure}

In this setup, SpikGPT successfully identified the missing cell type. Specifically, 97\% of ‘alpha cells’ in the query set were correctly rejected and labeled as ‘Unknown’, and the remaining were mostly assigned to ‘pancreatic polypeptide cells’ (PP), which are closely related endocrine cells (Fig~\ref{fig:hPancreasA}, Fig~\ref{fig:hPancreasB} and Fig~\ref{fig:hPancreasC}). As shown in the UMAP (Fig~\ref{fig:hPancreasA}), these ‘alpha cells’ formed a distinct cluster, further supporting their identity as a separate cell type.

This experiment demonstrates that SpikGPT possesses a unique and valuable ability to detect and abstain from classifying previously unseen cell types—an essential feature for real-world applications where reference datasets are often incomplete.










\section{Conclusion}
Our benchmarking results demonstrate that SpikeGPT achieves consistently high accuracy across diverse single-cell RNA sequencing (scRNA-seq) datasets, specifically including the retina and Human Lung Cell Atlas (HLCA) datasets. In particular, SpikeGPT outperforms or matches leading cell type annotation tools, such as scGPT, CCA, and scPred, when evaluated by macro precision, recall, and F1 score. This evidence signifies strong generalization capacity and inherent robustness against batch effects and biological heterogeneity. Notably, SpikeGPT performs competitively with foundation models like scGPT, while relying on a lightweight spiking transformer architecture, which confers distinct advantages in terms of improved efficiency.

Beyond conventional accuracy metrics, SpikeGPT exhibits a unique advantage in its ability to detect previously unseen cell types. This capability was empirically validated through a simulated experiment utilizing the hPancreas dataset. In this trial, a common cell type, the ‘alpha cells,’ was deliberately excluded from the training set. SpikeGPT successfully rejected 97\% of these alpha cells during inference by employing a confidence-based thresholding strategy. This finding emphatically highlights SpikeGPT’s ability to capture both precise cellular identities and meaningful uncertainty, which is a critical requirement for real-world applications involving novel or rare cell types.

Collectively, these results underscore the effectiveness of SpikeGPT as a scalable, and biologically aware model for automated cell annotation. SpikeGPT is demonstrably capable of both precise classification and robust novelty detection within complex transcriptomic landscapes.

\section{Code availability}
\url{https://github.com/warming151/spikeGPT.git}

\bibliographystyle{unsrtnat}
\bibliography{references}  






\end{document}